\documentclass[twocolumn,showkeys,preprintnumbers,amsmath,amssymb, floatfix]{revtex4}
\usepackage{graphicx}
\usepackage{dcolumn}
\usepackage{bm}
\usepackage{color}

\begin{document}

\title{Searching for the baryon-to-meson transition region with the MPD at NICA}

\author{Alejandro Ayala$^{1,2,3}$, Wolfgang Bietenholz$^{1}$, Eleazar Cuautle$^{1}$, Rodrigo Garc\'\i a Forment\'\i$^1$, Rodrigo Guzm\'an$^{1\footnote{Speaker}}$\\ \ \\}

\affiliation{%
$^1$Instituto de Ciencias Nucleares - Universidad Nacional Aut\'onoma de M\'exico, Apartado Postal 70-543, CdMx 04510, Mexico\\
$^2$Centre for Theoretical and Mathematical Physics and Department of Physics - University of Cape Town - Rondebosch 7700, South Africa\\
$^3$Departamento de F\'isica - Universidade Federal de Santa Maria - Santa Maria RS 97105-900, Brazil\\}%

\date{\today}
             
\begin{abstract}
In heavy-ion reactions, statistical models predict a rapid change in the baryon-to-meson ratio as a function of the collision energy. This change occurs when the hadronic medium transits from a baryon- to a 
meson-dominated gas. The transition is expected to take place at a temperature around 140 MeV and a baryon chemical potential around 420 MeV, corresponding to a center-of-mass collision energy of $\sqrt{s_{NN}}\simeq 8.2$ GeV. The energy range of the MPD experiment will be suitable for the exploration of this transition region. We present preliminary results of feasibility studies for the measurement of the transverse momentum spectra for mesons and baryons, using Monte Carlo data samples, in order to study the crossing point between these transverse momenta, as a function of the centrality and collision energy.
\end{abstract}

\keywords{heavy-ion collisions, transverse momentum spectra, baryon-to-meson transition}
                              
\maketitle


\section{\label{Intro}Introduction\protect}

One of the most active topics in heavy-ion physics is the study of the phase diagram of Quantum Chromodynamics, particularly the search for a critical endpoint (CEP), see {\it e.g.}\ Refs.\ ~\cite{STAR:2009sxc}. Experiments carried out in various facilities, such as the CERN-SPS~\cite{SPS} and the BNL-RHIC~\cite{RHIC}, have already explored some regions in parameter space. However, the energy region to be covered is so rich that so far we have only grasped the tip of the iceberg of this subject, and no CEP has been found yet.

Most of the current experiments cover the region of relatively high-energy. The Multi-Purpose Detector (MPD) is the flagship experiment of the Nuclotron-based Ion Collider fAcility (NICA), which is currently under construction at the Joint Institute for Nuclear Research (JINR) in Dubna, Russia. The MPD is designed to study a variety of ion collisions at high baryonic density in the energy range of $\sqrt{s_{NN}}=4$ to 11 GeV, which corresponds to a narrow low-energy region. This regime is particularly suitable for the investigation of the {\em baryon-to-meson transition}.

The thermal statistical model~\cite{SM}, which describes the chemical freeze-out, predicts  the abundance of mesons and baryons in collisions at a given energy~\cite{cleymans1}. This is realized by measuring the particle integrated yield ratios (for example $K^+/\pi^+$, $K^-/\pi^-$) and comparing to the predictions of the model to determine a temperature and a baryon chemical potential at freeze-out. The ratio $K^-/\pi^-$ shows a monotonic rise as the energy increases. For the ratio $K^+/\pi^+$, however, the model predicts a maximum at around $\sqrt{s_{NN}}=8.2$ GeV~\cite{cleymans1}. In this model, and for collisions at energies below this value, the contribution to the total entropy of the hadron gas is mainly carried by the baryon degrees of freedom. The meson contribution surpasses that of the baryons only for values above this energy. This can be interpreted as a transition from a baryon-to a meson-dominated gas and can be investigated by exploring the region where meson and baryon spectra cross, both, as function of transverse momenta and of collision energy.

The relevance of this transition region for the present study lies in the possibility to address the deeper question regarding the underlying production mechanisms. In addition, the analysis of the hadron abundances might provide some insights into the problem of the onset of deconfinement~\cite{SPS,onset}, and into properties of the QCD phase diagram~\cite{Hatta:2002sj}.


\section{\label{mpdexperiment} MPD Experiment }
The analysis is done in the framework of the software MPD\-Root~\cite{MPDroot}, simulating the transport of the particles through the subdetectors of the MPD. Our analysis employs mainly two detectors: the Time Projection Chamber (TPC) and the Time Of Flight  (TOF). The TPC surrounds the beam line and provides information about the particle trajectories, momentum and the energy loss ($dE/dx$). It consist of a gas mixture of 90\% Ar and 10\% CH$_4$ inside a cylindrical barrel of 3.4~m length and a radius of 1.4~m. The TPC is capable of tracking charged particles with transverse momentum larger than $50$ MeV/$c$ and covers a pseudorapidity range of $|\eta|\lesssim 1.2$. The TOF encloses the TPC and is composed of a gas mixture of 90\% C$_2$H$_2$F$_4$, 5\% SF$_6$ and 5\% i-C$_4$H$_{10}$. It makes use of the Multigap Resistive Plate Chambers technology, which provides information about the particle velocities with a time resolution around 50 ps. Both detectors are embedded inside a superconducting NbTi coil designed to provide a highly homogeneous magnetic field of 0.5 T in the beam line direction. Details on the design and other characteristics of the MPD can be found in Ref.~\cite{MPDstatus}.


\section{\label{data ana}Simulation and data analysis}

The  Monte Carlo simulation was carried out using the Ultra relativistic Quantum Molecular Dynamics (UrQMD) generator~\cite{UrQMD} at three energies: $\sqrt{s_{NN}}=7.7$ and 11.5 GeV for Au+Au collisions, and 9.2 GeV for Bi+Bi. For the reconstruction we employed GEANT3~\cite{GEANT3} to simulate the MPD response for the Au+Au collisions, and GEANT4~\cite{GEANT4} for the Bi+Bi case. Complete simulation and reconstruction analyses have already been reported by the MPD collaboration for particle spectra in Au+Au collisions at energies different from the ones considered in this work~\cite{MPDstatus}.

\subsection{\label{sel crit} Selection criteria}

For each of the collision events, only the primary charged particles are selected. This information is provided by the Monte Carlo simulation. The charged particles traveling through the TPC interact with the gas mixture, thus producing ionization clusters which then provide information on their trajectories.
Each  track, corresponding to one particle,  is reconstructed by considering a minimal number of hits (NofH): more than 13 hits for the Au+Au collisions, and more than 16 hits for the Bi+Bi case. The ratio $\chi^2/$NofH characterizes the quality of the reconstruction; we require it to be less than 8 for each track. The data is analyzed at midrapidity $|y|<0.5$ and in the transverse momentum range from 0.2 to 2 GeV/$c$, both in the TPC and the TOF, where the efficiency of the reconstruction remains on average above 80\%~\cite{MPDstatus}.

\subsection{\label{sel cent} Centrality selection}
\begin{figure}
\includegraphics[width=8.5cm,height=6cm]{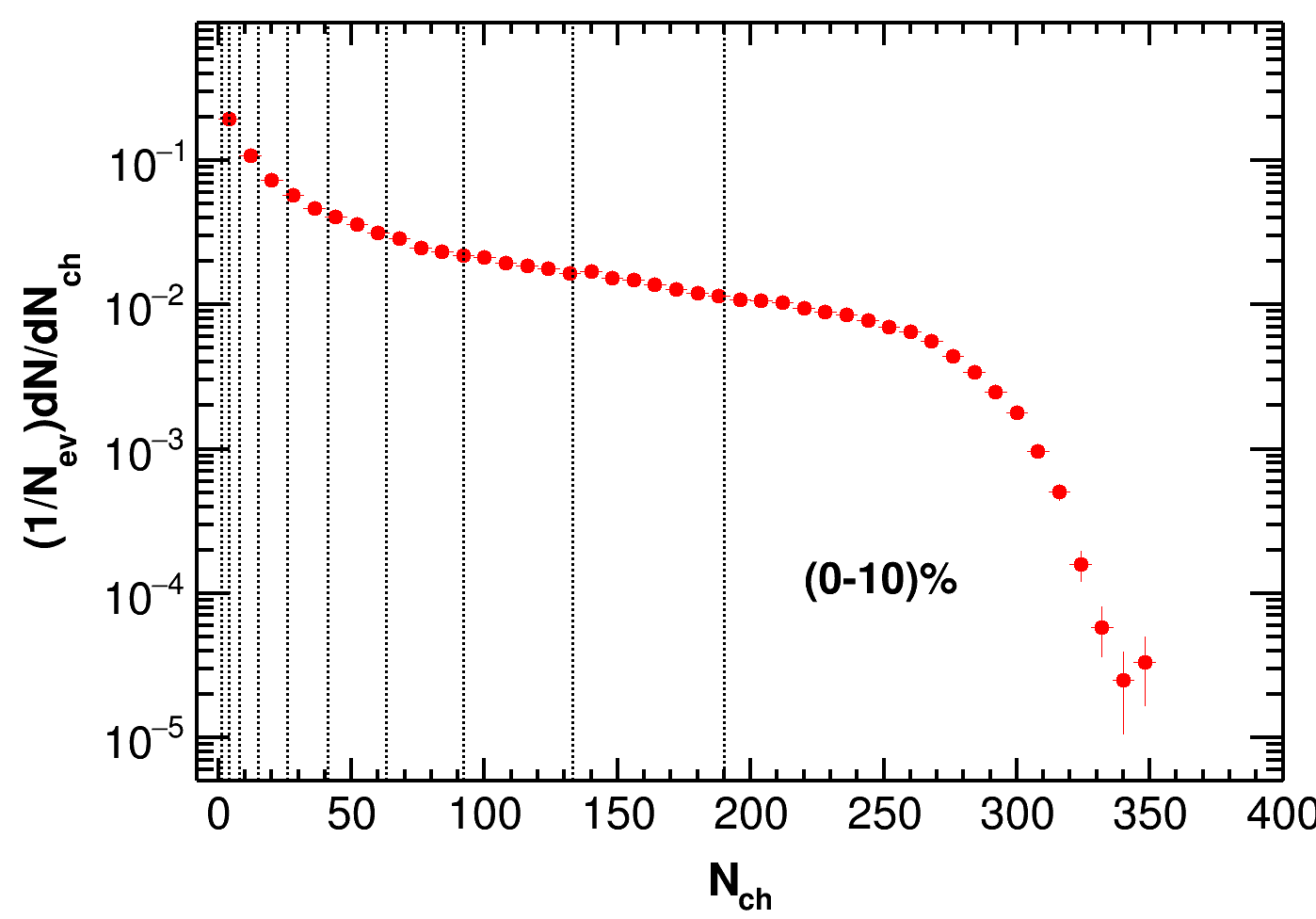}
\caption{\label{multi} Charged multiplicity distribution within $|\eta|<0.5$ for Au+Au collisions at $\sqrt{s_{NN}}=$ 11.5 GeV. The dashed vertical lines represent the centrality selection criteria employed in this work, 0-10\%, 10-20\%, etc.}
\end{figure}

The centrality classes are defined as sharp cuts in the multiplicity distribution. To define these cuts, we take a reference multiplicity, given by the primary charged particles at $|\eta|<0.5$, see Fig.~\ref{multi}. A fit is performed with the Bayesian inversion method ($\Gamma$-fit). The multiplicity distribution can be related to the geometric parameters of the collision in the Glauber Model~\cite{Glauber}, such as the impact parameter, number of participants etc., which are not directly measured. Given the simulated multiplicity and the values of the geometric parameters, we can compute fractions of the cross-section, and the centrality classes correspond to a given percentage of the cross-section~\cite{Klochkov:2017mpn,Zherebtsova:2018thp}.

We carried out this centrality selection for each of the collision energies, such that each event is included in one and only one of the 0-10\%, 10-20\%, 20-30\%, \dots , 70-80\% centrality classes. The results for even more peripheral collisions are not shown, because of the poor statistics. For further details on the selection process, see Ref.~\cite{Parfenov}.

\subsection{\label{ID} Particle identification}

\begin{figure}
\includegraphics[width=8.5cm,height=6cm]{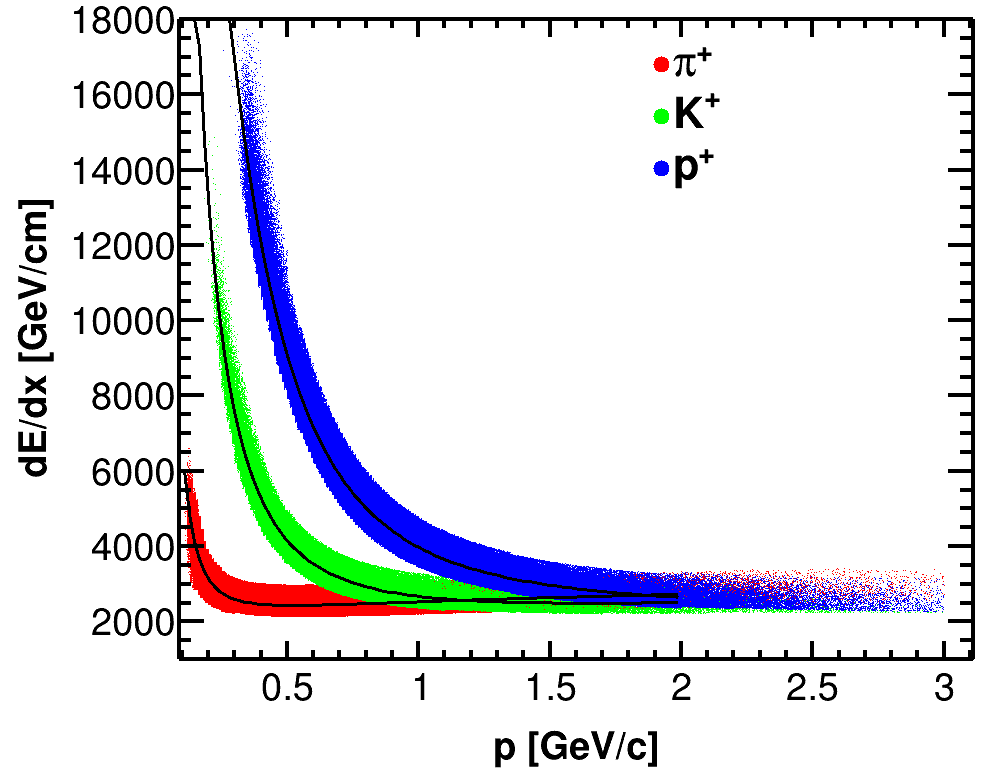}
\caption{\label{pid} Energy loss per distance vs.\ momentum for positively charged particles at $|y|<0.5$ for Au+Au collisions at $\sqrt{s_{NN}}=$11.5 GeV. Colored symbols represent values measured by the TPC and the black lines show the predicted energy loss according to the Allison-Cobb model ~\cite{allison}.}
\end{figure}

The identification is accomplished by referring to the energy loss per unit length, $dE/dx$, and the momentum measurements. Figure~\ref{pid} shows results for $dE/dx$ as a function of momentum for each of the particles identified by the TPC. In the Allison-Cobb model~\cite{allison}, the predicted energy loss is given by
\begin{eqnarray}
\biggl \langle \frac{dE}{dx} \biggr \rangle = \frac{p_1}{\beta^{p_4}}\biggl\{p_2+\beta^{p_4}-\ln\biggl[p_3+\biggl(\frac{1}{\beta\gamma}\biggl)^{p_5}\biggr]\biggr\}
\label{eq-allcobb},
\end{eqnarray}
where $\beta^2=p^2/(m^2c^2+p^2)$, $\gamma$ is the Lorentz factor, and $p_i$ are free fitting parameters. The fit for the Au+Au collisions at $\sqrt{s_{NN}}=$11.5~GeV is also shown in Fig.~\ref{pid} (solid black curves).

To perform the identification, each of the selected tracks is simultaneously assigned the probability of being either a proton, a kaon or a pion. This is carried out by a Bayesian approach ($n$-sigma method); for this analysis we chose 3 sigmas. Then, by applying a cut on the probability, the particle is identified as one of the particles of interest. From Fig.~\ref{pid} we observe that the measured distributions overlap at high momentum. The TPC alone can perform measurements particularly well for low momentum. The opposite is true for the TOF. Thus, a combination of these two detectors provides an overall improved performance. This is achieved by considering only the tracks from the TPC that are  matched with those from the TOF as identified particles.

\subsection{\label{eff} Efficiency and contamination}

\begin{figure}
\includegraphics[width=8.5cm,height=6cm]{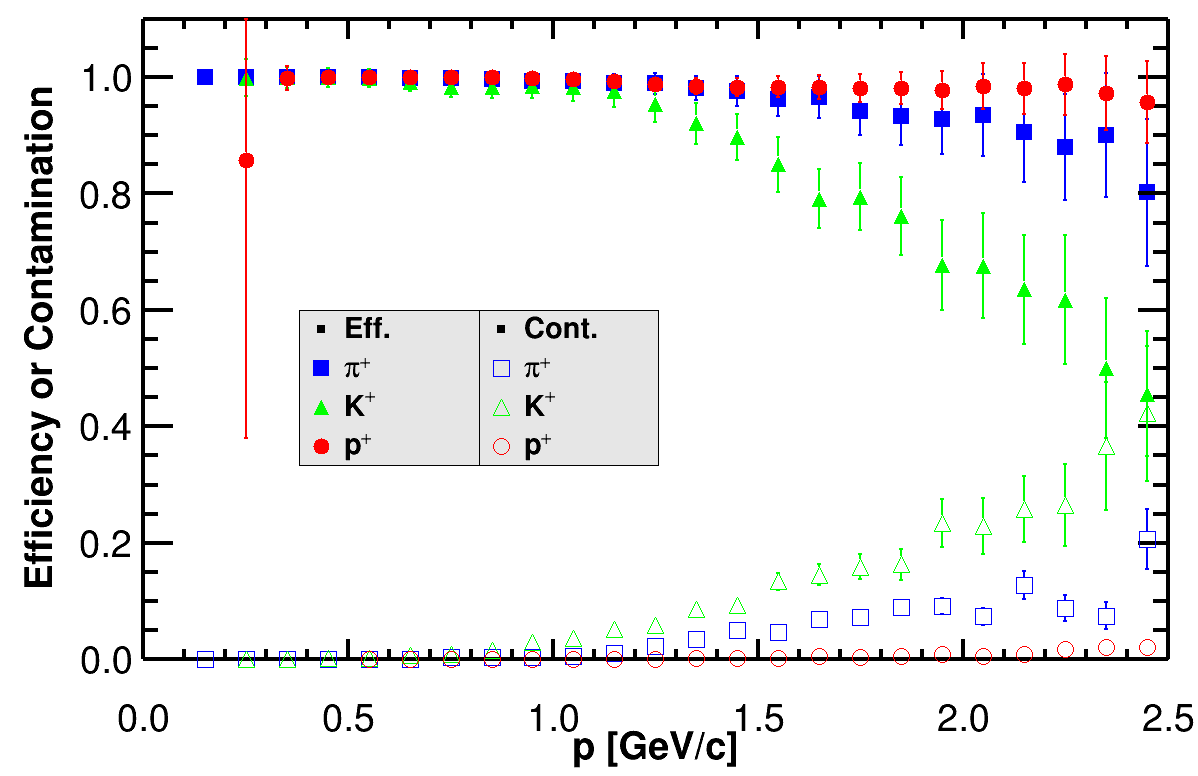}
\caption{\label{eff2} Efficiency and contamination for Au+Au collisions at $\sqrt{s_{NN}}=$ 7.7 GeV.}
\end{figure}

The {\em efficiency} is given by the ratio of the number of correctly reconstructed particles (after selection) divided by the total number of Monte Carlo generated particles. The {\em contamination} is obtained when the numerator of this ratio is substituted by the number of wrongly identified particles.
Figure~\ref{eff2} shows the efficiency and contamination for Au+Au collisions at $\sqrt{s_{NN}}=$ 7.7 GeV. Qualitatively similar results are obtained at other energies.

The efficiency decreases when $p$ increases. For instance, for the pions the efficiency remains more or less constant up to $p\simeq$ 1.7 GeV$/c$. At high momentum, due to the overlap of the distributions, the number of wrongly identified particles grows, and so does the contamination. The efficiency for the kaons decreases faster than for the pions and protons, reaching $\approx 80$\% at $p\simeq 1.7$ GeV$/c$. A mismatch is manifest when the tracks from the TPC are matched to the wrong tracks from the TOF; this usually happens at low momentum. This results in a reduced efficiency, particularly for the protons due to their larger mass.

The corrections by the efficiency and acceptance are not yet implemented at this stage. Therefore, most of the study has been conducted at the Monte Carlo level, using the reconstructed, uncorrected data with the purpose of qualitatively analyzing the gross features of the spectra.


\section{\label{results}Results and discussion}

\subsection{\label{pt} Transverse momentum distributions and crossing point}

\begin{figure*}
\includegraphics[width=17cm,height=12cm]{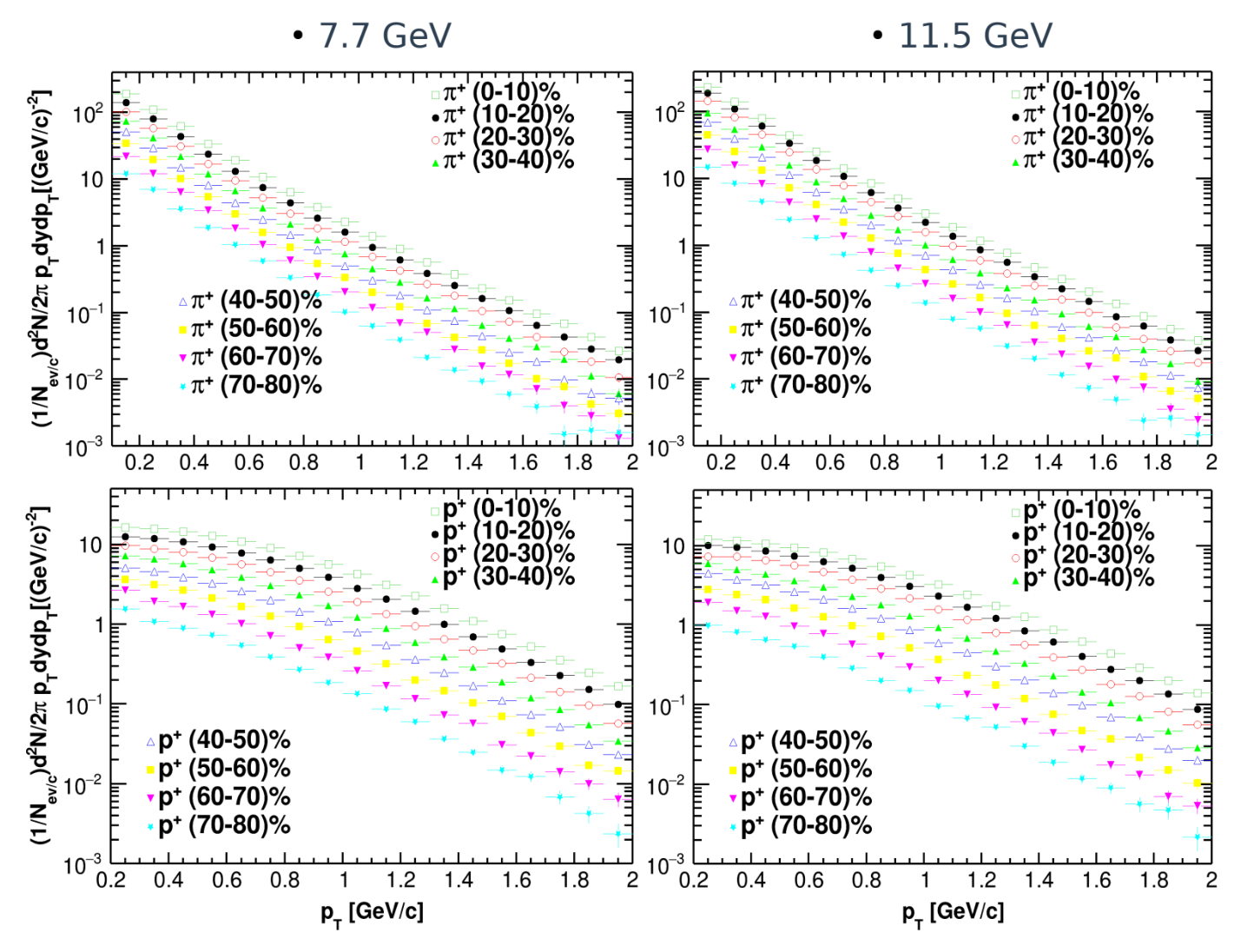}
\caption{\label{ppi}Transverse momentum distributions for positive charged pions (upper panels) and protons (lower panels), at $\sqrt{s_{NN}}=$ 7.7 GeV (left panels) and 11.5 GeV (right panels) for 8 centrality classes.}
\end{figure*}

Figure~\ref{ppi} shows the transverse momentum distributions for Au+Au collisions at $\sqrt{s_{NN}}=$ 7.7 and 11.5 GeV, and for all the centrality classes. The results for negative pions as well as for positive and negative kaons exhibit a similar behaviour. 

\begin{figure*}
\includegraphics[width=17cm,height=6cm]{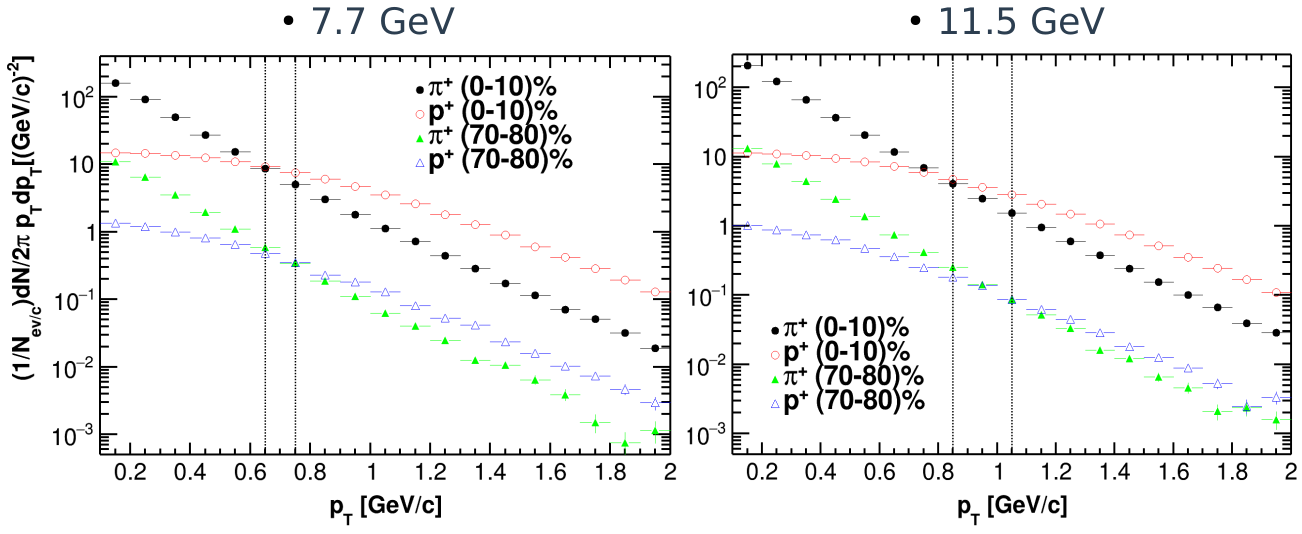}
\caption{\label{ppi cross}Comparison of the transverse momentum distributions for pions and protons for Au+Au collisions at  $\sqrt{s_{NN}}=$ 7.7 and 11.5 GeV. The vertical lines represent the approximate location of the crossing point for the two centrality classes.}
\end{figure*}

There is an interesting phenomenon which is manifest when we compare, at the same energy, the proton and pion distributions, for different centrality classes. Figure~\ref{ppi cross} shows that at low transverse momentum  $p_T$ the pion distributions reach higher values, which in turn means that they are more abundantly produced. However, for a given centrality and larger values of $p_T$, the situation is reversed and the proton abundance dominates. The value of the transverse momentum at which the distributions cross each other is hereby dubbed the {\em \lq\lq crossing point"}.  This crossing point moves from low to high $p_T$ when the energy increases, and centrality decreases. For instance for 0-10\% centrality, the crossing points are located at approximately $p_T \simeq  $ 0.65 and 0.85 GeV/$c$ for collision energies of $\sqrt{s_{NN}} = 7.7$~GeV and 11.5 GeV, respectively, whereas for the same energies but centrality 70-80\%, the crossing points are approximately at $p_T \simeq$ 0.75 and $p_T \simeq$ 1.05 GeV/$c$, respectively. The results for the Bi+Bi collisions at  $\sqrt{s_{NN}}=$ 9.2 GeV show the same behaviour, with the crossing point for the most central collisions located at approximately $p_T\simeq$ 0.55 GeV/$c$, whereas for the most peripheral collisions this occurs at $p_T\simeq$ 0.65 GeV/$c$. This means that the crossing point depends on the type of collision. To better study the evolution of the crossing point, either as a function of centrality or of the collision energy, the analysis should be performed considering the same type of collisions.

\subsection{\label{cp} Interpretation of the crossing point}

As the centrality of the collision decreases, we observe that the crossing point moves to a higher value of the transverse momentum, for both collision energies. Part of the observed effects are known to be due to the radial flow. In the expanding fireball, pions (with a small mass) are less affected by the transverse flow than the protons (which have a 6.7 times larger mass). Nevertheless, other effects, such as the particle production mechanism, may also play a role.

In the framework of the statistical model, one  could envision fitting the momentum distributions to extract information on the freeze-out parameters, including the strength of the radial flow. The residual features of the crossing point could then be attributed to properties such as the 
particle production mechanisms. Such a study could shed light on the possible change of the predominant particle production mechanism when the energy and centrality of the collision are varied. Further investigation in these directions is needed.

\subsection{\label{yields} Integrated particle yields}

The simplest way to study hadron abundances is by characterizing the particle integrated yields. To investigate the baryon-to-meson transition region, we considered the particle integrated $K^+/\pi^+$ ratio yields at the three energies of this study:  $\sqrt{s_{NN}}=$ 7.7 GeV and 11.5 GeV  for Au+Au,  and 9.2 GeV for Bi-Bi, in a limited  transverse momentum range ($0.4\leq p_T\leq 2$ GeV/$c$) for the three data sets at the Monte Carlo and reconstruction (but uncorrected) level, which provides some valuable insights. By taking the integrated yields ratio of the reconstructed to simulated data, we observed that the percentage of identified to generated $\pi^+$ is about $\simeq$ 66\% for both energies in the Au+Au case, and $\simeq$ 72\% in the Bi+Bi case. For the $K^+$ the corresponding rates are $\simeq$ 57\% and $\simeq$ 64\%. Thus, these $K^+/\pi^+$ integrated yield ratios cannot be compared to the reconstructed data unless corrected appropriately. The comparison can, however, be done at the Monte Carlo level.  The $K^+/\pi^+$ integrated yields ratio are $\simeq$ 0.229, $\simeq$  0.232 and $\simeq$ 0.205 for the data sets at $\sqrt{s_{NN}}=$ 7.7, 9.2 and 11.5 GeV, respectively. These values are compatible with the behaviour predicted by the statistical model, but still too preliminary to draw any definite conclusion.


\section{\label{sum}Summary and outlook}

The MPD is expected to be commissioned in the first part of the year 2024 and to start its run with Bi+Bi collisions at $\sqrt{s_{NN}}=$ 9.2 GeV. The study of the bulk properties of the matter produced in the reactions is one of the first experimental tasks. Meanwhile, Monte Carlo studies such as the ones reported here, together with comparisons with models, can provide useful insights for the tasks lying ahead.

In this work, we have discussed --- with a preliminary analysis --- the possibility to investigate the baryon-to-meson transition region using the MPD. We pointed out that the energy range of the detector is particularly well suited for this task and can provide new valuable information regarding the properties of the QCD phase diagram in the region of low energy and high baryon density.

Based on the study of the crossing point, we expect to be able to investigate some aspects related to flow and particle production mechanisms as a function of centrality and collision energy. The next step involves extracting the freeze-out parameters within the statistical model to better characterize the transition region from baryon-to-meson dominance.


\begin{acknowledgments}
We are very grateful for the permission to use the software of the MPD Collaboration.
We further acknowledge helpful comments and suggestions by Alexey Aparin, Alexander Mudrokh and Victor Riabov. Support for this work was received in part by UNAM-PAPIIT IG100322, and by the Consejo Nacional de Ciencia y Tecnolog\'ia (CONACYT), grant numbers A1-S-7655 and A1-S-16215.
\end{acknowledgments}



\begin{thebibliography}{}

\bibitem{STAR:2009sxc}
M.~A.~Stephanov, PoS \textbf{LAT2006} 024 (2006).
B.~I.~Abelev \textit{et al.} [STAR Collaboration],
Phys. Rev. C \textbf{81}, 024911 (2010).
K.~Fukushima and T.~Hatsuda,
Rept. Prog. Phys. \textbf{74}, 014001 (2011).
J.~M.~Pawlowski, AIP Conf. Proc. \textbf{1343} 75 (2011).
 
\bibitem{SPS}
M. Gazdzicki, Eur. Phys. J. Spec. Top. \textbf{229}, 3507 (2020).

\bibitem{RHIC}
L. Kumar, Mod. Phys. Lett. A \textbf{28}, 1330033 (2013).

\bibitem{SM}
M. Gazdzicki and M. Gorenstein, Acta Phys. Pol. B \textbf{30}, 2705 (1999).

\bibitem{cleymans1}
J. Cleymans \textit{et al.}, Phys. Lett. B \textbf{615}, 50 (2005).

\bibitem{onset}
M. Gazdzicki, M. Gorenstein and P. Seyboth, Acta Phys. Polon. B \textbf{42}, 307 (2011).

\bibitem{Hatta:2002sj}
Y. Hatta and T. Ikeda, Phys. Rev. D \textbf{67}, 014028 (2003).

\bibitem{MPDroot} http://mpdroot.jinr.ru/

\bibitem{cleymans2}
J. Cleymans and K. Redlich, Phys. Rev. Lett. \textbf{81}, 5284 (1998).

\bibitem{MPDstatus}
V. Abgaryan \textit{et al.} [MPD Collaboration], Eur. Phys. J. A \textbf{58}, 140 (2022).

\bibitem{UrQMD} S.~A.~Bass \textit{et al.}, Prog. Part. Nucl. Phys. \textbf{41}, 255
(1998). M. Bleicher \textit{et al.}, J. Phys. \textbf{G25} 1859 (1999).

\bibitem{GEANT3} R.~Brun \textit{et al.}, ebook: 10.17181/CERN.MUHF.DMJ1

\bibitem{GEANT4} http://geant4.cern.ch

\bibitem{Glauber}
M. Miller, K. Reygers, S. Sanders and P. Steinberg, Annu. Rev. Nucl. Part. Sci. \textbf{57}, 205 (2007).

\bibitem{allison}
M. Allison and H. Cobb, Annu. Rev. Nucl. Part. Sci. \textbf{30}, 253 (1980).

\bibitem{Klochkov:2017mpn}
V.~Klochkov and I.~Selyuzhenkov,
Acta Phys. Polon. Supp. \textbf{10}, 919 (2017).

\bibitem{Zherebtsova:2018thp}
E.~Zherebtsova, V.~Klochkov, I.~Selyuzhenkov, A.~Taranenko and E.~Kashirin,
EPJ Web Conf. \textbf{182}, 02132 (2018).

\bibitem{Parfenov}
P. Parfenov, D. Idrisov, V. B. Luong and A. Taranenko,
Particles \textbf{4}, 275 (2021).

\end{thebibliography}
\end{document}